\documentclass[aps,prd,twocolumn,showpacs,amsmath,nofootinbib]{revtex4-1} 
\usepackage{graphicx}
\usepackage{epsfig}

\usepackage{graphicx}
\usepackage{dcolumn}
\usepackage{bm}

\begin{document}

\title{Neutrino signals from ultracompact minihalos and constraints on the 
primordial curvature perturbation}

\author{Yupeng Yang$^{1,2,3}$} \email{yyp@chenwang.nju.edu.cn}
\author{Guilin Yang$^{1,3}$} \email{yanggl@chenwang.nju.edu.cn}
\author{Hongshi Zong$^{1,3,4}$} \email{zonghs@chenwang.nju.edu.cn}
\affiliation{ $^1$Department of Physics, Nanjing University, Nanjing, 210093, China\\
$^2$Department of Astronomy, Nanjing University, Nanjing, 210093, China\\
$^3$Joint Center for Particle, Nuclear Physics and Cosmology, Nanjing, 210093, China\\
$^4$ State Key Laboratory of Theoretical Physics, Institute of Theoretical Physics, CAS, Beijing, 100190, China}

\begin{abstract}
Compared with that of standard dark matter halos,
the density profile of the recently proposed new kind of dark matter structure
named ultracompact 
minihalos (UCMHs) is steeper and its formation time is earlier. If the dark matter is composed of weakly interactive 
massive particles (WIMP), the potential signals, e.g. neutrinos, 
from UCMHs due to the dark matter annihilation would be detected by IceCube/DeepCore or other 
detectors and such signals would have a very useful complementarity of $\gamma$-ray observations. 
On the other hand, the formation of UCMHs is related to primordial 
curvature perturbations on the smaller scales. So constraints on the 
abundance of UCMHs can be used to give a limit on the perturbations on
these scales. In previous works in the literature, the authors focused on the $\gamma$-ray signals 
from UCMHs due to dark matter annihilation. In this work, 
we investigate the neutrino signals from nearby UCMHs. 
Although no excess of neutrino signals the dark matter 
annihilation has been observed, the constraints on the abundance of UCMHs can 
be obtained and these constraints can be 
translated into the limit on the primordial curvature perturbations on small scales. 

\end{abstract}


\maketitle
\section {Introduction}

In the earlier epoch, primordial black holes (PBHs)
would be formed if there were large density perturbations 
($\delta \rho/\rho \ge 0.3$) \cite{9704251}. On the other hand, 
theoretical research and many observations have shown that the 
present cosmic structures comes from the earlier density perturbations 
which have smaller amplitude $\delta \rho /\rho \sim 10^{-5}$ \cite{wmap7}. 
Recently, Ricotti and Gould proposed that a new kind of structure 
named ultracompact minihalos (UCMHs) would be formed if the density perturbations 
satisfy the conditions: $3 \times 10^{-4} \lesssim \delta \rho/\rho \lesssim 0.3$ \cite{0908.0735}. 
Compared with classical dark matter halos, the formation time of these objects is earlier and the density profile
is steeper. So 
it is expected that these new interesting structures would have an effect 
on the cosmological evolution, such as reionization and recombination, 
due to the dark matter annihilation within them if the 
dark matter is composed of weakly interacting massive particles 
(WIMP) \cite{dz,yyp_prd,yyp_epjp}, such as neutralinos. According to the theory, dark matter 
can annihilate into the standard particles, such as photons($\gamma$), 
electrons and positrons($e^+,e^-$), 
and UCMHs would become one kind of possible high energy astrophysical 
sources, such as $\gamma$-ray sources \cite{scott_prl,1006.4970,yyp_jcap}.
The other kind of important product of dark matter annihilation 
is the neutrino. 
Although this kind of particle 
has nearly no effect on the process of reionization or recombination, 
dark matter halos would be the potential neutrino sources \cite{0905.2075,0906.4364,1002.0197}. 
According to the present theory and observations,
\footnote{The new constraint on the number of relativistic species coming from 
the WMAP-9 year data is $N_{\mathrm{eff}} = 3.26\pm0.35$ \cite{wmap9}.} 
there are three types of neutrino, $\nu_{e}, \nu_{\mu}, \nu_{\tau}$, and 
they can convert into one another during propagation. The virtue of neutrino 
detection is that the neutrinos can propagate nearly without attenuation, so the 
orientation of the sources can be confirmed easily. Among them, 
muons ($\mu$) can be produced by 
the charged flux interaction when they propagate through the medium and they can be 
detected by way of the Cherenkov radiation which is produced when 
{\bf $\mu$} propagates through a medium such as water. So, $\nu_{\mu}$ is the main target 
particle of the conventional detection. Generally, the production of gamma-ray flux goes 
with the neutrino signals when the dark matter annihilates. Therefore, the 
research on neutrino signals would be a good complementarity of gamma-ray 
detection. In this paper, different from Refs. \cite{scott_prl,1006.4970,yyp_jcap} 
where the authors have focused on the gamma-ray flux, we will investigate the potential 
neutrino signals from the UCMHs due to dark matter annihilation. 

Besides being potential high energy astrophysical sources, UCMHs 
are related to the primordial power spectrum of density (curvature) 
perturbations on small scales due to their earlier formation time. 
On larger scales ($k \sim 10^{-4} - 1 \mathrm{Mpc}^{-1}$), 
the primordial power spectrum of curvature perturbations  
can be constrained by the CMB, Lyman-$\alpha$ and large scale 
structure observations \cite{wmap7_2,1105.4887,lyman,lar_stru}. 
All of these observations show a nearly scale-invariant 
spectrum of primordial perturbations with an amplitude
$\mathcal{P}_\mathcal{R}(k) \sim 10^{-9}$. 
\footnote{The new results from the WMAP-9 year data show that 
there is a tilt in the primordial spectrum \cite{wmap9}.} 
On smaller scales 
($k \sim 1 - 10^{20} \mathrm{Mpc}^{-1}$), the main 
constraints come from the PBHs, $\mathcal{P}_\mathcal{R}(k) \sim 10^{-2}$ \cite{pbh_pps}. 
If the UCMHs are formed during the earlier epoch, 
these objects could also be used to obtain the constraints on smaller scales. 
In Refs.~\cite{1006.4970,1110.2484}, the authors first used the $\gamma$-ray observations 
to obtain the constraints on the abundance of UCMHs for different masses 
and then obtained the limit on the primordial curvature perturbations for the 
corresponding scales, $\mathcal{P}_\mathcal{R}(k)\sim 10^{-7}-10^{-6}$ 
for $k \sim 5 - 10^{8} \mathrm{Mpc}^{-1}$. The complementary constraints 
can also be obtained from the effect of astrometric 
microlensing produced by UCMHs \cite{1202.1284}. 
In this work, we will use the neutrino observations to obtain the 
constraints on the primordial power spectrum of curvature perturbations for 
the small scales. 

This paper is organized as follows. The neutrino signals from UCMHs due to 
dark matter annihilation are studied in Sec. II. 
In Sec. III, the constraints on the abundance of UCMHs are obtained 
using the neutrino observations. Then, using these results, we obtain the constraints on
the primordial power spectrum of curvature perturbations for 
the small scales. The conclusions are presented in Sec. IV.

\section {Neutrino signals from UCMHs due to the dark matter annihilation} 

\subsection{The basic quality of UCMHs}
After the seeds of UCMHs are formed, the mass will increase through the 
radial infall. During the radiation dominated epoch, the increase 
of the mass is slow until after the redshift of equality of radiation and 
matter. The variation of the mass of UCMHs with the redshift can be written as 
\cite{fillmore,scott_prl}

\begin{equation}
\label{Mh}
M_\mathrm{UCMHs}(z) = M_i \left(\frac{1 + z_\mathrm{eq}}{1+z}\right),
\end{equation}
where $M_i$ is the mass within the perturbations scale at the redshift $z_{eq}$.
The density profile of UCMHs can be obtained 
from the simulation \cite{fillmore,bert}, 
$\rho \sim r^{-9/4}$, and the specific form is   

\begin{equation}
\rho(r,z) = \frac{3f_\chi M_\mathrm{UCMHs}(z)}{16\pi R(z)^\frac{3}{4}r^\frac{9}{4}},
\end{equation}
where ${R(z)}=0.019\left(\frac{1000}{z+1}\right)\left(\frac{M(z)}
{\mathrm{M}_\odot}\right)^\frac{1}{3} \mathrm{pc}$ and 
$f_{\chi} = \frac{\Omega_{DM}}{\Omega_b+\Omega_{DM}} = 0.83$~\citep{wmap7} 
is the dark matter fraction which describes that only dark matter 
content collapsed to form the UCMHs in the beginning. Due to the structure 
formation effect, the mass of UCMHs will stop increasing soon and 
in this work we assume the corresponding redshift is $z \sim 10$ \cite{scott_prl,
1110.2484}. The radius of UCMHs is $R_{(z=10)} = 0.01 M_i^\frac{1}{3}$ and 
$R \sim $1 kpc for $M_i = 10^6 M_\odot$. So, following the previous works
~\cite{scott_prl,1006.4970,1110.2484}, in this paper
we also treat the UCMHs as point sources.

\subsection{Neutrino signals from nearby UCMHs}

The annihilation rate of dark matter is proportional to the 
number density squared, $\Gamma \sim n^2\langle\sigma v\rangle 
= \rho^2\langle\sigma v\rangle/m^2$, so the UCMHs have very significant 
effect on the cosmological evolution \cite{yyp_prd,yyp_jcap,yyp_epjp}. 
In Refs. \cite{scott_prl,1110.2484,yyp_jcap}, the authors 
have investigated the $\gamma$-ray flux from the nearby and extra-Galactic UCMHs. 
Besides these high energy 
photons, the neutrinos would be produced together with the $\gamma$-ray 
in the process of dark matter annihilation. 
There are three flavors of neutrinos and their anti-neutrinos, 
$\nu_e (\bar \nu_e),\nu_\tau (\bar \nu_\tau),\nu_\mu (\bar \nu_\mu)$, and
they can convert into one another due to 
the vacuum oscillation effect. The muon neutrinos ($\nu_\mu$) can convert to 
muons ($\mu$) during their propagation due to the charged current interaction with 
the matter. These muon signals can be detected by the detectors on Earth 
through, for example, the Cherenkov light. Some muons would be produced in the detectors and 
some are produced before arriving at the detectors. 
In this paper, we consider these two kinds of neutrino signals which are named
contained and upward events, respectively.
 
\begin{figure}
\epsfig{file=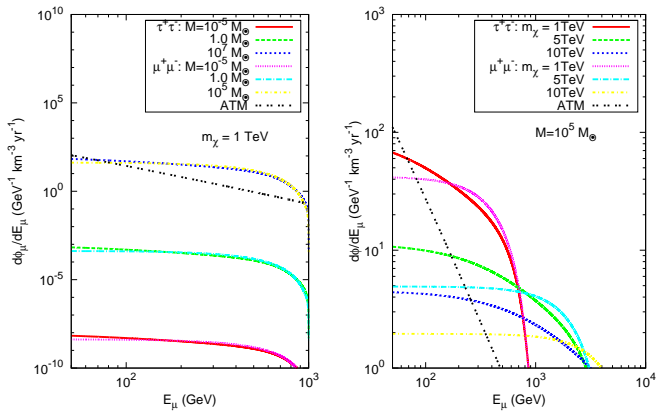,width=0.5\textwidth}
\caption{The contained events of muons from UCMHs due to dark matter 
annihilation. Two channels are shown: $\tau^+ \tau^-$, $\mu^+ \mu^-$.  
Left: The dark matter mass is fixed $m_\chi$ = 1 TeV, and the masses of UCMHs 
are $\mathrm{M_{UCMHs}} = 10^{-5}, 1.0$ and $ 10^{5} \mathrm{M_\odot}$ respectively.
Right: The dark matter masses are $\mathrm{m_\chi}$ = 1, 5 and 10 TeV and the mass of UCMHs 
is $\mathrm{M_{UCMHs}} = 10^{5}$. For these results, the cross section of dark matter 
annihilation has been set as $\langle \sigma v \rangle = 3.0 \times 10^{-26} \mathrm{cm^3 s^{-2}}$. 
In both figures, the angle-averaged muon's flux ($\theta_{\mathrm{max}} = 5^{\circ}$)
for the atmospheric neutrino is also shown (ATM).}
\label{fig:fig1}
\end{figure}

The muons produced in the detector through the charged current interactions 
are called "contained events". Following Ref.~\cite{1009.2068},
this kind of flux can be written as

\begin{eqnarray}
\frac{d\phi_\mu}{dE_\mu} = \frac{N_A\rho}{2}
\int^{m}_{E_\mu} dE_{\nu}\left(\frac{d\phi_\nu}{dE_\nu}\right) 
\left(\frac{d\sigma^{p}_{\nu}(E_\nu,E_\mu)}
{dE_{\mu}}+(p \to n)\right) \nonumber\\ +(\nu \to \bar\nu),
\end{eqnarray}
where $N_A = 6.022 \times 10^{23}$ is Avogadro's number and $\rho$ is 
the density of the medium. $\frac{d\sigma^{p,n}_{\nu,\bar \nu}}{dE_{\mu}}$ are 
the scattering cross sections of neutrinos and antineutrinos off 
protons and neutrons. $\frac{d\phi_\nu}{dE_\nu}$ is 
the differential flux of neutrinos from UCMHs due to the dark matter annihilation 

\begin{eqnarray}
\frac{d\phi_\nu}{dE_\nu} = \frac{1}{8\pi}\frac{dN_\nu}{dE_\nu}\frac{\langle\sigma v\rangle}{m_{\chi}^2 d^2}
\int \rho^2(r)4\pi r^2dr,
\end{eqnarray}
where $\frac{dN_\nu}{dE_\nu}$ is the neutrino number per dark matter annihilation and 
can be obtained from the public code DarkSUSY \cite{darksusy}. In this paper, 
we consider the ratio between the neutrino flavors as 1:1:1. Here, 
we have treated UCMHs as point sources and $d$ is the distance of UCMHs from Earth. 
The flux of the contained events is shown in the Fig.~\ref{fig:fig1}. Two channels 
($\tau^+ \tau^-$, $\mu^+ \mu^-$) are shown and the distance is $d = 10$ kpc. For 
the mass of UCMHs, we have chosen three values: $\mathrm{M_\mathrm{UCMHs}}$ = $10^{-5}, 1.0,$ and $10^{5} 
M_\odot$. 

The main background of the neutrino signals is the atmospheric neutrino flux. 
For the spectrum of these flux, we use the form \cite{neu_for_1, neu_for_2}
\begin{eqnarray}
\frac{d\phi_\nu}{dE_{\nu} d\Omega}&&=N_0E_\nu^{-\gamma-1}  \nonumber \\
 &&\times \left(\frac{a}{1+bE_{\nu}cos\theta}
+\frac{c}{1+eE_{\nu}cos\theta}\right),
\label{eq5}
\end{eqnarray}
where $\gamma = 1.74, a = 0.018, b = 0.024, c = 0.0069, e = 0.00139, N_0 = 1.95 \times 
10^{17}$ for neutrinos and $ 1.35 \times 10^{17}$ for antineutrinos. 
For the neutrino detector, e.g. the IceCube, the angular resolution is $\theta = 1^{\circ} \sim 3^{\circ}$ 
for the energy ranges considered by us~\cite{resolu_neu}. 
On the other hand, the angle between the muon
and the neutrino in the neutrino-nucleon scattering should also be considered. 
Therefore, we set the angle $\theta_{\mathrm{max}} = 5^{\circ}$ and it is 
enough for our work. In addition, it should be noted 
that the more detailed analysis might lead to an even
better signal to background ratio than indicated in this work, especially at
higher energies, where smaller opening angles might be possible.

From Fig.~\ref{fig:fig1}, one can see that the bigger the masses of UCMHs are, 
the larger the final muon flux is. 
For $\mathrm{M_{UCMHs}} = 10^{5} \mathrm{M_\odot}$, the muon flux exceeds 
the ATM flux which becomes lower for the higher energy. 
For a fixed mass of UCMHs, the muon flux will be larger for bigger 
dark matter mass and higher muon energy. 
Because the flux of ATM decreases as the energy increases, the neutrino 
signals from UCMHs due to dark matter annihilation would be detected 
possibly at higher energy for the large dark matter mass.

\begin{figure}
\epsfig{file=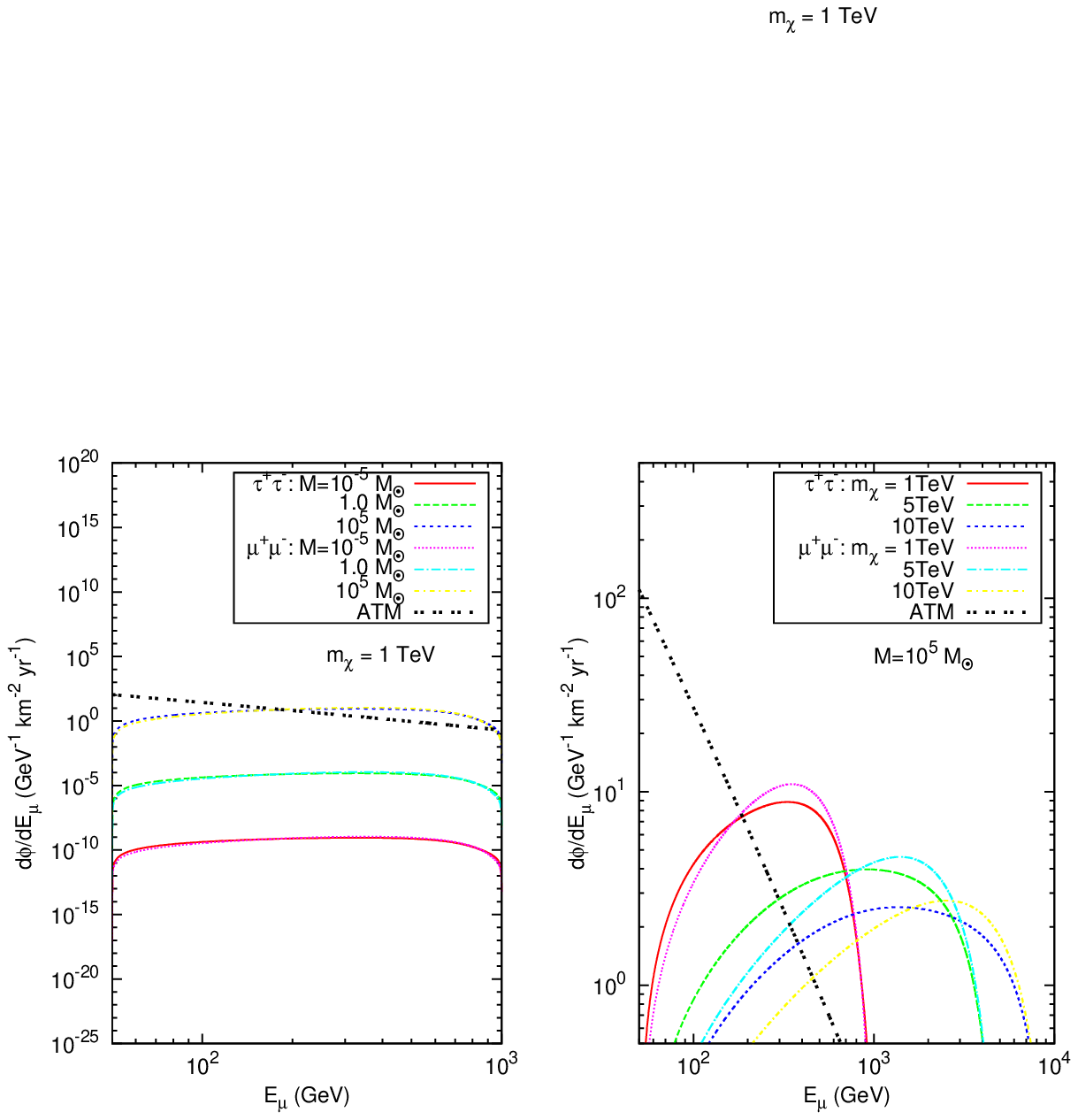,width=0.5\textwidth}
\caption{The upward events of muons from UCMHs due to dark matter 
annihilation. Here, we have chosen the threshold of the detector to be $E_{\mu}^{th} = 
50 \mathrm{GeV}$. The other related parameters used here are the same as those in 
Fig.~\ref{fig:fig1}.}
\label{fig:fig2}
\end{figure}

The muon neutrinos can also convert into muons before arriving at 
the detector. The flux for this case is called `upward events' 

\begin{eqnarray}
&&\frac{d\phi_\mu}{dE_\mu} = \frac{N_A\rho}{2} \int dE_\nu \left(\frac{d\phi_\nu}{dE_\nu}\right)
\nonumber \\ &&\times \left(\frac{d\sigma^{p}_{\nu}
(E_\nu,E_\mu)}{dE_\mu} +(p \to n)\right)R(E_\mu)+(\nu \to \bar \nu), 
\end{eqnarray}
where $R(E_\mu)$ is the distance at which muons can propagate in matter 
until its energy is below the threshold of the detector $E_{\mu}^{th}$ \cite{1002.0197,
0912.0513} and its form is  
$R(E_\mu) = \frac{1}{\rho \beta}ln\frac{\alpha+\beta E_{\mu}}{\alpha
+\beta 
E_{\mu}^{th}}$ with $\rho$ being the density of the medium, $\alpha \sim 10^{-3}\mathrm{GeV 
cm^2/g}$ and $\beta \sim 10^{-6} \mathrm{cm^2/g}$. The upward events from the UCMHs 
are shown in Fig.~\ref{fig:fig2}. 
From this figure one can see that the muon flux for the fixed dark matter 
mass (left) is comparable with the contained events. The muons with 
energy below the threshold of the detector cannot arrive at the detector, so the 
signals are cut off near the energy $E \sim 50$ GeV. 
A common character of the contained and upward events is that the flux of muons from 
the UCMHs due to the dark matter annihilation exceeds the flux of the atmospheric muons for 
larger masses of UCMHs and dark matter. For these cases, the UCMHs can possibly be 
detected. 

\subsection{Comparison with other observations}
As we mention in the introduction, previous works focused on the $\gamma$-ray 
signals from the UCMHs. In this section, we will discuss these aspects and it can be 
seen that although the gamma-ray detection is much more attractive, the 
neutrino detection is still competitive in some cases. 
In Ref.~\cite{scott_prl}, the authors studied the $\gamma$-ray signals 
from the nearby UCMHs ($d_\mathrm{{UCMHs}} = 100 pc$) formed during 
three phase transitions in the early Universe (electroweak symmetry breaking, QCD confinement and
$e^+e^-$ annihilation). They found that for some cases the integrated flux above 100 MeV exceeds
the thresholds of EGRET or Fermi. For example, for the $b \bar b$ 
channel, the $\gamma$-ray signals from the UCMHs formed during the 
$e^+e^-$ annihilation period exceed the threshold of EGRET and Fermi for 
the dark matter mass $m_\chi \sim 10 - 1000 \mathrm{GeV}$. For the lepton channel, $\mu^+\mu^-$, 
the flux is lower but still above the threshold value. 
So, if there are UCMHs within this distance, 
they would be observed by Fermi. In order to have a comparison with 
our results, we find that the integrated gamma-ray flux for the $\mu^+\mu^-$ channel 
in Ref.~\cite{scott_prl} will be rescaled by a factor $\sim 10^{-2}$ for $d \sim 10 \mathrm{kpc}$ and 
$M_{\mathrm{UCMHs,0}} \sim 10^5 M_{\odot}$. 
As a result, the flux will be under the threshold of the EGRET or Fermi detectors 
for the larger dark matter mass, e.g. $m_{\chi} \sim 1 \mathrm{TeV}$. 
The situations are different for the neutrino signals. As shown in the left plot in Fig.~\ref{fig:fig1}, 
the neutrino signals from the UCMHs are still larger than the ATM. From the right plot it can also be seen 
that the situation is much better for the larger dark matter mass. 
But it should be noticed that for the lighter dark matter e.g. $m_{\chi} \sim 100 \mathrm{GeV}$, 
due to the stronger flux of ATM, the gamma-ray detection are much more competitive than the neutrino case. 
Other interesting targets which are similar to the UCMHs are the dwarf spheroidals which 
are confirmed to be dark matter dominated. In Ref.~\cite{1001.4531}, the potential 
$\gamma$-ray flux is studied using the Fermi-LAT detector. No 
obvious excess has been observed and these nondetection results can be transformed 
into the limits on the dark matter parameters, e.g. dark matter mass ($m_{\chi}$) and 
the thermal averaged annihilation cross section ($\langle\sigma v\rangle$). 
The limits are different for different dark matter models and different 
dwarf spheroidals. The neutrino signals have also been studied using the IceCube detector and there is 
no excess of signals~\cite{1111.2738}. The limits on the dark matter parameters can also be 
derived from these results. It can be seen that for the neutrino signals the constraints are weaker. On the other hand, the $\gamma$-ray and the neutrino 
signals are channel dependent. The constraints on the dark matter from the $\gamma$-ray detection 
are stronger for $b \bar b$ or $W^+W^-$ channels. One main reason for this is that the integrated 
number of these channels is larger than that of the lepton channels. The situation is 
different for the neutrino detection and the limits are stronger for the 
$\mu^+\mu^-$ and $\tau^+\tau^-$ channels. For example, 
the authors of~\cite{0912.0513} studied the neutrino signals from nearby dwarfs. 
They found that compared with the current 
gamma-ray detectors, for the $\mu^+\mu^-$ or $\nu_{\mu} \bar \nu_{\mu}$ channels, the IceCube 
neutrino detector will be competitive, especially for large dark dark matter mass, $m_{\chi} \gtrsim 7 \mathrm{TeV}$. 
Similar results can also be seen from Refs.~\cite{1111.2738,1108.3546}. For the neutrino detection, 
the main contamination for the source signals is from the atmospheric neutrino flux. 
From Eq.~\ref{eq5} it can be seen that these flux decreases with increasing energy as $\sim E_{\nu}^{-3}$. 
On the other hand, the muon flux from the UCMHs does not decrease as much. 
So the neutrino signals would be detected for higher energy which 
corresponds to the larger dark matter mass or UCMHs. On top of this, 
with the improvement of the angular resolution of neutrino telescopes, 
it is expected that the limits on dark matter will be much better. 
Therefore, compared with the gamma-ray detectors, 
the neutrino detection has its own advantage, especially for  
heavier dark matter (e.g. $m_{\chi} \gtrsim 1 \mathrm{TeV}$) 
and lepton channels (e.g. $\mu^+\mu^-$ or $\nu_{\mu} \bar\nu_{\mu}$). 
The other most important point is that they will be a very useful complementarity 
of $\gamma$-ray observations for seeking dark matter.

\section{Constraints on the primordial curvature perturbation}

The primordial curvature perturbation is very important 
for modern cosmology and its amplitude can be limited from different 
observations. The main constraints come from the cosmic microwave background (CMB) 
$\mathcal{P}_\mathcal{R}(k) \sim 10^{-9}$ \cite{wmap7_2}. 
These constraints carry the information of primordial curvature 
perturbations which correspond to the scales 
$k \sim 10^{-4}-1\mathrm{Mpc}^{-1}$. For smaller scales, the main constraints 
come from the PBHs in spite of no observations of these objects, 
$\mathcal{P}_\mathcal{R}(k) \sim 10^{-2}$ for 
$k\sim10^{-2}-10^{19} \mathrm{Mpc^{-1}}$~\cite{pbh_pps}. The UCMHs 
provide another way of constraining the primordial curvature perturbations 
on smaller scales. Due to the steep density profile of UCMHs, it is excepted 
that the $\gamma$-ray flux would be produced from dark matter annihilation. 
In order to be consistent with the present observations, such as $Fermi$, 
the abundance of UCMHs must be constrained. In Refs.~\cite{1006.4970,1110.2484}, 
the authors used the points sensitivity of $Fermi$ to obtain the final constraints, 
$\mathcal{P}_\mathcal{R}(k) \sim 10^{-6} - 10^{-8}$ for 
$k\sim 5-10^{7.5} \mathrm{Mpc^{-1}}$. 
As mentioned in the introduction, the neutrinos are usually 
produced together with the $\gamma$-ray flux. Therefore, these signals would be a 
significant complementarity of $\gamma$-ray detection. Different from the previous works, 
we study the potential neutrino signals 
from UCMHs due to dark matter annihilation. Because no excess of neutrino signals 
has been observed as compared with the ATM, 
we can obtain the conservative constraints on the abundance of UCMHs and 
they can be used to obtain the limit on the primordial curvature 
perturbations. The methods used by us are mainly from Refs. \cite{1110.2484,1202.1284,1006.4970} 
and here we only show the main description of the calculations. 
 
The cosmological mass fraction at horizon entry, $\beta(M_H)$, 
which then forms the UCMHs, is related to the present fraction of UCMHs 
\cite{1110.2484,1202.1284}, 

\begin{equation}
\Omega_\mathrm{UCMHs} = \Omega_\mathrm{DM}
\frac{\mathrm{M}_\mathrm{UCMHs}(z=0)}{\mathrm{M}_
\mathrm{UCMHs}(z_{eq})}\beta(M_\mathrm{H}),
\end{equation}
where $\Omega_\mathrm{UCMHs}$ and $\Omega_\mathrm{DM}$ are the fractions of 
UCMHs and DM, respectively, 
$\mathrm{M_{UCMHs}}(z=0, z_{eq})$ is the mass of UCMHs at present and the redshift of equality of matter 
and radiation. If the initial perturbations are Gaussian, the 
present fraction of UCMHs can be written in the form~\cite{1006.4970}

\begin{eqnarray}
\Omega_\mathrm{UCMHs} = &&\frac{2\Omega_\mathrm{DM}}{\sqrt{2\pi}\sigma_{H}(R)}
\frac{\mathrm{M}(z=0)}{\mathrm{M}(z_{eq})} \times  \nonumber \\  
&&\int^{\delta_{\mathrm{max}}}_{\delta_{\mathrm{min}}}\mathrm{exp}\left(-\frac{\delta^2_H(R)}{2\sigma^2_H(R)}\right)
d\delta_H(R),
\end{eqnarray}
where $\delta_{\mathrm{max}}$ and $\delta_{\mathrm{min}}$ 
are the maximal and minimal values of density perturbations required 
for the formation of UCMHs. Both of them depend on the redshift \cite{1110.2484} 
and in this work, for simplicity, we choose these values as 
$\delta_{\mathrm{max}} = 0.3$ and $\delta_{\mathrm{min}} = 10^{-3}$, respectively.
$\sigma_H(R)$ is related to the curvature perturbation as~\cite{1110.2484,1202.1284}

\begin{equation}
\sigma^2_{H}(R) = \frac{1}{9}\int^{\infty}_{0}x^3W^2(x)\mathcal{P}_{\mathcal{R}}
(x/R)T^2(x/\sqrt{3})dx,
\end{equation} 
where $W(x) = 3x^{-3}(sinx-xcosx)$ is the Fourier transform of the top-hat 
windows function with $x \equiv kR$. $T$ is the transfer function describing 
the evolution of perturbations. For more detailed discussions one can see the 
appendixes in Refs. \cite{1110.2484,1202.1284}. 
The fraction of UCMHs can be defined as ~\cite{1006.4970} 

\begin{equation}
\frac{\Omega_\mathrm{UCMHs}}{\Omega_\mathrm{DM}} = 
\frac{\mathrm{M_{UCMHs}}(z=0)}{\mathrm{M_{DM}}(r < d_{obs})},
\end{equation}
where $M_{DM}(r < d_{obs})$ is the mass within the radius $d_{obs}$ which 
is the distance on which the neutrino signals from UCMHs would be observed 
by the detector. In this work, we use the NFW profile for the dark matter halo 
of the Milky Way and assume that the abundance of UCMHs is the same everywhere.
\footnote{For different density profiles of dark matter halos, 
the final constraints would be different~\cite{yyp_decay}. On the other 
hand, similar to the PBH case~\cite{pbhs_clusters_1,pbhs_clusters_2}, 
UCMH clusters would be formed during the earlier epoch.} Much more 
accurate definition is given in Refs.~\cite{frac_acc_1, frac_acc_2}.

The detection of neutrino signals is more difficult than that of the gamma rays 
due to the weak interactions between neutrinos and other particles. 
Moreover, the ATM would contaminate the target signals. In order to reduce 
this effect, Earth itself is usually used as the shield, which means 
that the good targets should lie on the other side of Earth 
compared to the detectors. Therefore, for the fixed target one should 
consider which detector is good. In this work, we assume that the UCMHs 
studied by us just lie in the appropriate direction. 
At present or in the near future, there are several detectors which can be used to 
search neutrino signals, such as IceCube/DeepCore and KM3Net.
\footnote{http://en.wikipedia.org/wiki/List\_of\_neutrino\_experiments.} 

\begin{figure}
\epsfig{file=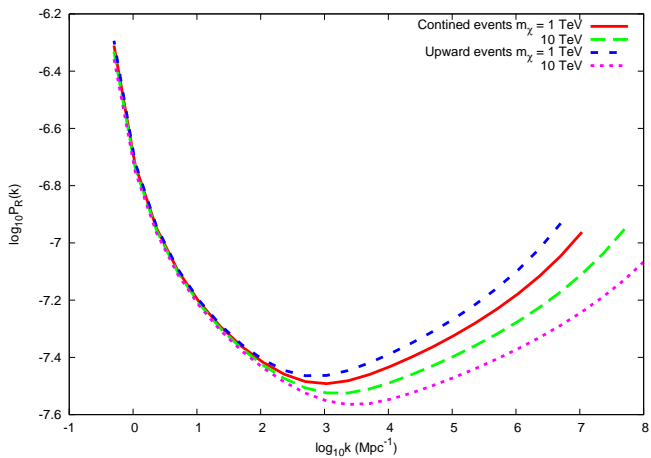,width=0.5\textwidth}
\caption{The constraints on the primordial curvature perturbations on the scales 
$k \sim 1 - 10^{8} \mathrm{Mpc^{-1}}$. 
Here we have set 10 years exposure time for IceCube/DeepCore and $2 \sigma$ 
statistical significance for ATM. Constraints for both upward and contained events 
are shown. Two kinds of dark matter mass are considered: 
$m_{\chi}$ = 1 TeV and 10 TeV. }
\label{fig:fig3}
\end{figure}

Considering the contamination of ATM, for an exposure time such as 
ten years, the minimal number of neutrinos from UCMHs 
which satisfies, e.g. $2\sigma$ statistic significance, can be obtained through \cite{9702037}

\begin{equation}
T_{obs} = \sigma^2\frac{N_\mathrm{ATM} + N_\mathrm{UCMHs}}{N_\mathrm{UCMHs}^2},
\end{equation}
where $N_\mathrm{UCMHs}$ is the number of neutrinos from UCMHs due to 
dark matter annihilation.  
It can be obtained 
by integration of Eqs.(3) and (6) 

\begin{equation}
N_\mathrm{UCMHs} = \int^{E_{max}}_{E_{\mu}^{th}}\frac{d\phi_{\mu}}{dE_{\mu}}F_\mathrm{eff}(E_\mu)dE_\mu,
\end{equation}
where $F_{\mathrm{eff}}(E_\mu)$ correspond to the 
effective volume $V_{\mathrm{eff}}$ and effective area $A_{\mathrm{eff}}$ 
of the detector for the contained and upward events, respectively. 
For IceCube/DeepCore, we accept that the energy independent 
effect volume is $V_\mathrm{eff} = 0.04 \mathrm{km^3}$ 
and the angle-averaged muon effective area 
is $A_{\mathrm{eff}} = 1 \mathrm{km}^2$. From Figs. 1 and 2, it can be seen that the neutrino 
fluxes for the two channels $\tau^+ \tau^-$ and $\mu^+ \mu^-$ are slightly different, but 
the final integrated number is nearly the same. So, in this work, for simplicity we only consider one channel. The final constraints on the primordial curvature perturbation 
are shown in Fig.~\ref{fig:fig3}. For this plot, we have chosen two 
values of the dark matter mass, $m_{\chi}$ = 1 and 10 TeV, and 
the exposure time of the detector is ten years. 
From this figure, one can see that on the larger scales, $k \lesssim 10^3 \mathrm{Mpc}^{-1}$, 
the constraints are nearly the same for different events and dark matter mass. 
The results are obviously different on larger scales 
$10^3  \lesssim k \lesssim 10^{8} \mathrm{Mpc}^{-1} $. 
Stronger constraints come from the upward events and larger dark matter mass. 
For the contained events, the strongest constraint is $\mathcal{P}_\mathcal{R}(k) \sim 10^{-7.5}$ for 
$k \sim 10^3 \mathrm{Mpc^{-1}}$, while for the upward events, the strongest 
constraint is $\mathcal{P}_\mathcal{R}(k) \sim 10^{-7.6}$ for 
$k \sim 10^{3.5} \mathrm{Mpc^{-1}}$.

\section{Conclusion}

Due to the steep density profile of UCMHs, it is expected that the 
dark matter annihilation rate is very strong within these objects. The $\gamma$-ray flux 
as the product of dark matter annihilation has been discussed 
by several authors. On the other hand, neutrinos would be produced 
together with the gamma rays. In this work, we have studied this kind of potential 
signal including upward events and contained events. 
We found that for the fixed distance of UCMHs ($d$ = 10 kpc) 
the final muon flux would exceed the flux for the ATM case for the large mass 
of UCMHs (e.g.$M_\mathrm{UCMHs} \sim 10^{5}M_{\odot}$) 
and dark matter (e.g.$m_{\chi} \sim 10^4$GeV). Because the 
flux of ATM decreases as the energy increases, the detection of high energy 
neutrinos is very significant for the indirect search of dark matter. 
On the other hand, compared with the classical dark matter halos, the formation time 
of UCMHs is earlier, so 
the abundance of UCMHs can be used to constrain the primordial curvature 
perturbations on small scales, which are not achieved through CMB observations. 
In the previous works in the literature, the authors obtained the constraints using the $\gamma$-ray 
observations or through the microlensing effect. 
The neutrino signals as a useful complementarity of gamma rays 
can also be used to achieve this goal. In this work, comparing with the neutrino background from 
the atmosphere, we obtain the constraints on the neutrino numbers from UCMHs 
due to dark matter annihilation and use these results to obtain the limit 
on the primordial curvature perturbations on the scales $k \sim 1 - 10^9 \mathrm{Mpc^{-1}}$. 
For the contained events, the strongest limit is $\mathcal{P}_\mathcal{R}(k) \sim 10^{-7.5}$ for 
$k \sim 10^3 \mathrm{Mpc^{-1}}$, and for the upward events, 
the strongest limit is $\mathcal{P}_\mathcal{R}(k) \sim 10^{-7.6}$ for 
$k \sim 10^{3.5} \mathrm{Mpc^{-1}}$. In the previous works~\cite{1006.4970,1110.2484}, 
the limits on the primordial curvature perturbations for the smaller scales are 
also obtained for the nondetection of gamma-ray signals. The strongest limits 
are $\mathcal{P}_\mathcal{R}(k) \sim 10^{-7}$ and our results are slightly stronger than these limits.
However, it should be noted that different formation times of UCMHs 
also affect the final results (Fig.5 in Ref.\cite{1110.2484}). 
For future detectors, such as KM3Net, due to
their larger effective volume (or area) and lower threshold, 
it is expected that the constraints on the primordial curvature perturbations 
on small scales will be stronger.

\section{Acknowledgments}
We thank Weimin Sun for improving the manuscript. Yupeng Yang
thanks Xiangyu Wang, Lei Feng and Qiang Yuan for useful discussion. 
This work is supported in part by the National Natural Science
Foundation of China (under Grants No. 10935001, 11275097 and No.
11075075) and the Research Fund for the Doctoral Program of Higher
Education (under Grant No 2012009111002)

\newcommand\PR[3]{~Phys. Rept.{\bf ~#1}, #2~(#3)}
\newcommand\NJP[3]{~New J.Phys.{\bf ~#1}, #2~(#3)}
\newcommand\PRD[3]{~Phys. Rev. D{\bf ~#1}, #2~(#3)}
\newcommand\APJ[3]{~Astrophys. J.{\bf ~#1}, #2~ (#3)}
\newcommand\PRL[3]{~Phys. Rev. Lett.{\bf ~#1}, #2~(#3)}
\newcommand\APJS[3]{~Astron. J. Suppl.{\bf ~#1}, #2~(#3)}
\newcommand\EPJP[3]{~Eur. Phy. J. Plus{\bf ~#1}, #2~(#3)}
\newcommand\Nature[3]{~Nature{\bf ~#1}, #2~(#3)}
\newcommand\JCAP[3]{~JCAP{\bf ~#1}, #2~(#3)}
\newcommand\APJL[3]{~Astrophys. J. Lett.{\bf ~#1}, #2~ (#3)}
\newcommand\MNRAS[3]{~MNRAS{\bf ~#1}, #2~(#3)}
\newcommand\PLB[3]{~Phys. Lett. B{\bf ~#1}, #2~(#3)}
\newcommand\AP[3]{~Astropart. Phys.{\bf ~#1}, #2~(#3)}
\newcommand\ARNPS[3]{~Ann. Rev. Nucl. Part. Sci.{\bf ~#1}, #2~(#3)} 
\newcommand\JPCS[3]{~J. Phys.: Conf. Ser.{\bf ~#1}, #2~(#3)}


\begin{thebibliography}{99}
\bibitem{9704251} A. M. Green, A. R. Liddle, \PRD{56}{6166}{1997} 
\bibitem{wmap7} E. Komatsu et al, \APJS{192}{18}{2011}
\bibitem{0908.0735} M. Ricotti, A. Gould, \APJ{707}{979}{2009}, arXiv:0908.0735
\bibitem{dz} D. Zhang, \MNRAS{418}{1850}{2011}, arXiv:1011.1935
\bibitem{yyp_prd} Y. Yang, X. Huang, X. Chen, H. Zong, \PRD{84}{043506}{2011}
\bibitem{yyp_epjp} Y. Yang, X. Chen, T. Lu, H. Zong, \EPJP{126}{123}{2011}
\bibitem{scott_prl} P. Scott, S. Sivertsson, \PRL{103}{211301}{2009}
\bibitem{yyp_jcap} Y. Yang, L. Feng, X. Huang, X. Chen, T. Lu, H. Zong, \JCAP{12}{020}{2011} 
\bibitem{1006.4970} A. S. Josan, A. M. Green, \PRD{82}{083527}{2010}, arXiv:1006.4970
\bibitem{0905.2075} J. Hisano, K. Nakayama, M. J. S. Yang, \PLB{678}{101}{2009}, arXiv:0905.2075
\bibitem{0906.4364} A. E. Erkoca, M. H. Reno, I. Sarcevic, \PRD{80}{043514}{2009}, 
arXiv: 0906.4364
\bibitem{1002.0197} Q. Yuan, P. Yin, X. Bi, X. Zhang, S. Zhu, \PRD{82}{023506}{2010}, 
arXiv:1002.0197
\bibitem{wmap9} G. Hinshaw et al, arXiv:1212.5226 
\bibitem{wmap7_2} D. Larson et al, \APJS{192}{16}{2011}, arXiv:1001.4635
\bibitem{1105.4887} R. Hlozek et al, arXiv:1105.4887
\bibitem{lyman} S. Bird, H. V. Peiris, M. Viel, L. Verde, \MNRAS{413}{1717}{2011}, arXiv:1010.1519
\bibitem{lar_stru} J. L. Tinker et al, \APJ{745}{16}{2012}
\bibitem{pbh_pps} A. S. Josan, A. M. Green, K. A. Malik, \PRD{79}{103520}{2009}, arXiv:0903.3184
\bibitem{1110.2484} T. Bringmann, P. Scott, Y. Akrami, \PRD{85}{125027}{2012}, arXiv:1110.2484
\bibitem{1202.1284} F. Li, A. L. Erickcek, N. M. Law, \PRD{86}{043519}{2012}, arXiv:1202.1284
\bibitem{fillmore} J. A. Fillmore, P. Goldreich, \APJ{281}{1}{1984}
\bibitem{bert} E. Bertschinger, \APJS{58}{39}{1985}
\bibitem{1009.2068} A. E. Erkoca, M. H. Reno, I. Sarcevic, \PRD{80}{043514}{2009}, arXiv:1009.2068
\bibitem{darksusy} http://www.physto.se/~edsjo/darksusy/
\bibitem{neu_for_1} T.K. Gaisser, M. Honda, \ARNPS{52}{153}{2002}
\bibitem{neu_for_2} M. Honda, T. Kajita, K. Kasahara, S. Midorikawa, T. Sanuki, \PRD{75}{043006}{2007}
\bibitem{resolu_neu} J. Dumm, H Landsman for the IceCube Collaboration, \JPCS{60}{334}{2007} 
\bibitem{0912.0513} P. Sandick, D. Spolyar, M. Buckley, K. Freese, D. Hooper, 
\PRD{81}{083506}{2010}, arXiv:0912.0513
\bibitem{1001.4531} The Fermi-LAT Collaboration, \APJ{712}{147}{2010}
\bibitem{1111.2738} The IceCube Collaboration, arXiv:1111.2738
\bibitem{1108.3546} The Fermi-LAT Collaboration, \PRL{107}{241302}{2011}, arXiv:1108.3546
\bibitem{frac_acc_1} T. Bringmann, P. Scott and Y. Akrami, \PRD{85}{125027}{2012}
\bibitem{frac_acc_2} S. Shandera, A. L. Erickcek, P. Scott and J. Y. Galarza, arXiv:1211.7361
\bibitem{9702037} L. Bergstrom, J. Edsjo, M. Kamionkowski, \AP{7}{147}{1997}
\bibitem{yyp_decay} Y. Yang, Guilin Yang, Hongshi Zong, arXiv:1210.1409
\bibitem{pbhs_clusters_1} J. R. Chisholm, \PRD{73}{083504}{2006}, astro-ph/0509141
\bibitem{pbhs_clusters_2} J. R. Chisholm, \PRD{84}{124031}{2011}, arXiv:1110.4402
\end{thebibliography}
\end{document}